\documentclass[twocolumn,floatfix]{revtex4}
	\usepackage{amssymb}
	\usepackage{epsfig}
	\usepackage{amsmath}
	\usepackage{graphicx}
	
\begin{document}

\newcommand{\comment}[1]{}

\newcommand{\eps}{\epsilon}
\newcommand{\e}{{\rm e}}
\newcommand{\dm}{{\rm d}m}
\newcommand{\dx}{{\rm d}x}
\newcommand{\ds}{{\rm d}s}
\newcommand{\dt}{{\rm d}t}
\newcommand{\du}{{\rm d}u}
\newcommand{\ex}[1]{{\rm E}[#1]}
\newcommand{\poi}{{\rm Poisson}}

\newcommand{\aobs}{\alpha_{\rm obs}}

	\title{Accuracy and Scaling Phenomena in Internet Mapping}
	\author{Aaron Clauset$^*$ and Cristopher Moore$^{*,\dagger}$}
	\affiliation{$^*$Computer Science Department and $^\dagger$Department of Physics and Astronomy, University of New Mexico, Albuquerque NM 87131 \\ 
	{\tt (aaron,moore)@cs.unm.edu}}
	\date{\today}

\begin{abstract}
A great deal of effort has been spent measuring topological features of the Internet.  However, it was recently argued that sampling based on taking paths or {\em traceroutes} through the network from a small number of sources introduces a fundamental bias in the observed degree distribution.  We examine this bias analytically and experimentally.  For Erd\H{o}s-R{\'e}nyi random graphs with mean degree $c$, we show analytically that traceroute sampling gives an observed degree distribution $P(k) \sim k^{-1}$ for $k \lesssim c$, even though the underlying degree distribution is Poisson. For graphs whose degree distributions have power-law tails $P(k) \sim k^{-\alpha}$, traceroute sampling from a small number of sources can significantly underestimate the value of $\alpha$ when the graph has a large excess (i.e., many more edges than vertices).  We find that in order to obtain a good estimate of $\alpha$ it is necessary to use a number of sources which grows linearly in the average degree of the underlying graph.  Based on these observations we comment on the accuracy of the published values of $\alpha$ for the Internet.
\end{abstract}
\maketitle

The Internet is a canonical complex network, and a great deal of effort has been spent measuring its topology.  However, unlike the Web where a page's outgoing links are directly visible, we cannot typically ask a router who its neighbors are.  As a result, studies have sought to infer the topology of the Internet by aggregating paths or {\em traceroutes} through the network, typically from a small number of sources to a large number of destinations~\cite{Pansiot, Govindan, IMP, skitter, Opte}, routing decisions like those imbedded in Border Gateway Protocol (BGP) routing tables~\cite{BGP, Amini, Oregon}, or both~\cite{Faloutsos, Rocketfuel, LookingGlass}.  Although such methods are known to be noisy~\cite{Amini, Chen, Barford, Teixeira}, they strongly suggest that the Internet has a power-law degree distribution at both the router and domain levels.

However, Lakhina et al.~\cite{Lakhina} recently argued that traceroute-based sampling introduces a fundamental bias in topological inferences, since the probability that an edge appears within an efficient route decreases with its distance from the source. They showed empirically that traceroutes from a single source cause Erd\H{o}s-R\'enyi random graphs $G(n,p)$, whose underlying distribution is Poisson~\cite{gnp}, to appear to have a power law degree distribution $P(k) \sim k^{-1}$. Here, we prove this evocative result analytically by modeling the growth of a spanning tree on $G(n,p)$ using differential equations.

Although it is widely accepted that the Internet, unlike $G(n,p)$, has a power-law degree distribution  $P(k) \sim k^{-\alpha}$ with $2 < \alpha < 3$ \cite{Faloutsos}, we may reasonably ask whether
traceroute sampling accurately estimates the exponent $\alpha$. Petermann and de los Rios~\cite{paolo} and Dall'Asta~\cite{DallAsta} considered this question, and found that because low-degree vertices are undersampled relative to high-degree ones, the observed value of $\alpha$ is lower than the true exponent of the underlying graph.  We explore this idea further, and find that single-source traceroute sampling only gives a good estimate of $\alpha$ when the underlying graph has a small excess, i.e., has average degree close to $2$ and is close to a tree.  As the average degree grows, so does the extent to which traceroute sampling underestimates $\alpha$.  

Since single-source traceroutes can signficantly underestimate $\alpha$, we then turn to the question of how many sources are required to obtain an accurate estimate of $\alpha$.  We find that the number of sources needed increases linearly with the average degree.  We conclude with some discussion of whether the published values of $\alpha$ for the Internet are accurate, and how to tell experimentally whether more sources are needed.

\medskip {\em Traceroute spanning trees: analytical results.}
The set of traceroutes from a single source can be modeled as a spanning tree~\cite{mult}.  If we assume that Internet routing protocols approximate shortest paths, this spanning tree is built breadth-first from the source.  In fact, the results of this section apply to spanning trees built in a variety of ways, as we will see below.

We can think of the spanning tree as built step-by-step by an algorithm that explores the graph.  At each step, every vertex in the graph is labeled \mbox{{\em reached}}, \mbox{{\em pending}}, or \mbox{{\em unknown}}.  Pending vertices are the leaves of the current tree; reached vertices are interior vertices; and unknown vertices are those not yet connected.  We initialize the process by labeling the source vertex pending, and all other vertices unknown.  
Then the growth of the spanning tree is given by the following pseudocode:
\begin{quote}
\begin{tabbing}
{\tt while}\=\ there are pending vertices: \\
\>choose a pending vertex $v$ \\
\>label $v$ reached \\
\>for \=every unknown neighbor $u$ of $v$, \\
\> \> label $u$ pending.
\end{tabbing}
\end{quote}
The type of spanning tree is determined by how we choose the pending vertex $v$.  Storing vertices in a queue and taking them in FIFO (first-in, first-out) order gives a breadth-first tree of shortest paths; if we like we can break ties randomly between vertices of the same age in the queue, which is equivalent to adding a small noise term to the length of each edge as in~\cite{Lakhina}.  Storing pending vertices on a stack and taking them in LIFO (last-in, first-out) order builds a depth-first tree.  Finally, choosing $v$ uniformly at random from the pending vertices gives a ``random-first'' tree.

\begin{figure} [htbp]
\begin{center}
\includegraphics[scale=0.45]{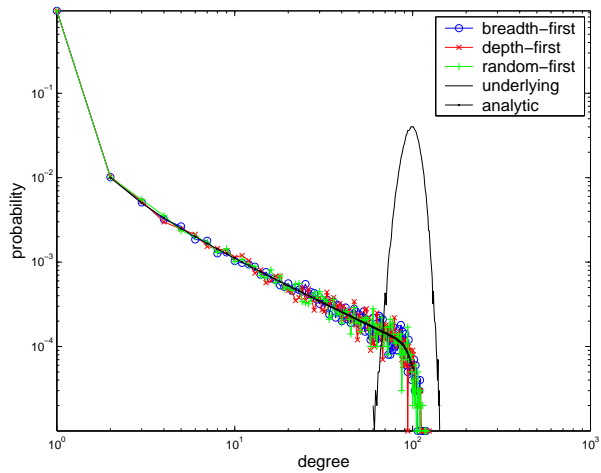}
\caption{Sampled degree distributions from breadth-first, depth-first and random-first spanning trees on a random graph of size $n=10^5$ and average degree $c=100$, and our analytic results (black dots). For comparison, the black line shows the Poisson degree distribution of the underlying graph. Note the power-law behavior of the apparent degree distribution $P(k) \sim k^{-1}$, which extends up to a cutoff at $k \sim c$. }
\label{fig:graph}
\end{center}
\end{figure}

Surprisingly, while these three processes build different trees, and traverse them in different orders, they all yield the same degree distribution when $n$ is large. To illustrate this, Fig.~\ref{fig:graph} shows the degree distributions for each type of \comment{breadth-first, depth-first, and random-first} spanning tree for a random graph $G(n,p=c/n)$ where $n=10^{5}$ and $c=100$.  The three degree distributions are indistinguishable, and all agree with the analytic results derived below.

We now show analytically that building spanning trees in Erd\H{o}s-R\'enyi random graphs $G(n,p=c/n)$ using any of the processes described above gives rise to an apparent power law degree distribution $P(k) \sim k^{-1}$ for $k \lesssim c$.  To model the progress of the {\tt while} loop described above, let $S(T)$ and $U(T)$ denote the number of pending and unknown vertices at step $T$ respectively.  The expected changes in these variables at each step are 
\begin{eqnarray}
\ex{U(T+1)-U(T)} & = & - p U(T) \nonumber \\
\ex{S(T+1)-S(T)} & = & p U(T) - 1 \label{eq:diff}
\end{eqnarray}
Here the $p U(T)$ terms come from the fact that a given unknown vertex $u$ is connected to the chosen pending vertex $v$ with probability $p$, in which case we change its label from unknown to pending; the $-1$ term comes from the fact that we also change $v$'s label from pending to reached.  Moreover, these equations apply no matter how we choose $v$; whether $v$ is the ``oldest'' vertex (breadth-first), the ``youngest'' one (depth-first), or a random one (random-first).  Since edges in $G(n,p)$ are independent, 
the events that $v$ is connected to each unknown vertex $u$ are independent and occur with probability $p$.  

Writing $t=T/n$, $s(t) = S(tn)/n$ and $u(tn)=U(t)/n$, the difference equations~\eqref{eq:diff} become the following system of differential equations,
\begin{equation}
\frac{\du}{\dt} = -c u \enspace , \quad
\frac{\ds}{\dt} = c u - 1 
\label{eq:sys}
\end{equation}
With the initial conditions $u(0) = 1$ and $s(0) = 0$, the solution to~\eqref{eq:sys} is
\begin{equation}
\label{eq:sol}
u(t) = \e^{-ct} , \quad s(t) = 1 - t - \e^{-ct} \enspace .
\end{equation}
The algorithm ends at the smallest positive root $t_0$ of $s(t) = 0$; using Lambert's function $W$, defined as $W(x) = y$ where $y \e^y = x$, we can write
\begin{equation}
\label{eq:t0}
 t_0 = 1 + \frac{1}{c} W(-c\e^{-c}) \enspace .
\end{equation}
Note that $t_0$ is the fraction of vertices which are reached at the end of the process, and this is simply the size of the giant component of $G(n,c/n)$.

Now, we wish to calculate the degree distribution $P(k)$ of this tree.  The degree of each vertex $v$ is the number of its previously unknown neighbors, plus one for the edge by which it became attached (except for the root).  Now, if $v$ is chosen at time $t$, in the limit $n \to \infty$ the probability it has $k$ unknown neighbors is given by the Poisson distribution with mean $m = cu(t)$,
$\poi(m,k) = \e^{-m} m^k / k!$.
Averaging over all the vertices in the tree and ignoring $o(1)$ terms gives
\[ P(k+1) = \frac{1}{t_0} \int_0^{t_0} \dt \,\poi(cu(t),k) \enspace . \]
It is helpful to change the variable of integration to $m$.  Since $m = c \e^{-ct}$ we have $\dm = -cm \,\dt$, and 
\begin{eqnarray}
 P(k+1) & = & \frac{1}{t_0} \int_{c(1-t_0)}^c \dm \,\frac{\poi(m,k)}{c m} \nonumber \\
 & \approx & \int_{c \e^{-c}}^c \dm \,\frac{\poi(m,k)}{cm} \nonumber \\
 & = & \frac{1}{c k!} \int_{c \e^{-c}}^c \dm \,\e^{-m} m^{k-1} \enspace .
 \label{eq:p1}
\label{eq:int}
\end{eqnarray}
Here in the second line we use the fact that $t_0 \approx 1 - \e^{-c}$ when $c$ is large (i.e., the giant component encompasses almost all of the graph).

The integral in~\eqref{eq:int} is given by the difference between two incomplete Gamma functions.  However, since the integrand is peaked at $m=k-1$ and falls off exponentially for larger $m$, for $k \lesssim c$ it coincides almost exactly with the full Gamma function $\Gamma(k)$.  Specifically, for any $c > 0$ we have
\[ \int_0^{c \e^{-c}} \dm \,\e^{-m} m^{k-1} < c \e^{-c} \]
and, if $k - 1 = c (1-\eps)$ for $\eps > 0$, then
\begin{eqnarray*}
 \int_c^\infty \dm \,\e^{-m} m^{k-1} 
 & = & \e^{-c} c^{k-1} \int_0^\infty \dx \,\e^{-x} (1+x/c)^{k-1} \\
 & < & \e^{-c} c^{k-1} \int_0^\infty \dx \,\e^{-x} \e^{x(k-1)/c} \\
 & = & \frac{\e^{-c} c^{k-1}}{\eps} 
 < \frac{\e^{-(k-1)} (k-1)^{k-1}}{\eps} \\
& \approx & \frac{\Gamma(k)}{\eps \sqrt{2 \pi (k-1)}}
 \end{eqnarray*}
\comment{
and, if $k - 1 = c - \Delta$ for $\Delta > 0$, then
\begin{eqnarray*}
 \int_c^\infty \dm \,\e^{-m} m^{k-1} 
 & = & \e^{-c} c^{k-1} \int_0^\infty \dx \,\e^{-x} (1+x/c)^{k-1} \\
 & < & \e^{-c} c^{k-1} \int_0^\infty \dx \,\e^{-x} \e^{x(k-1)/c} \\
 & = & \frac{\e^{-c} c^k}{\Delta} \\
 & < & \frac{c}{\Delta} \,\e^{k-1} (k-1)^{k-1} \\
 & \approx & \frac{c}{\Delta} \frac{\Gamma(k)}{\sqrt{2 \pi k}} \enspace .
\end{eqnarray*}
}
This is $o(\Gamma(k))$ if $\eps \gtrsim 1/\sqrt{k}$, i.e., if $k < c - c^\alpha$ for some $\alpha > 1/2$.  In that case we have
\begin{equation}
P(k+1) = (1-o(1)) \frac{\Gamma(k)}{c k!} 
\sim \frac{1}{ck}
\end{equation}
giving a power law $k^{-1}$ up to $k \sim c$.

We note that this derivation can be made mathematically rigorous, at least for constant $c$.  Wormald~\cite{Wormald} showed, under fairly generic conditions, that discrete stochastic processes like this one are well-modeled by the corresponding differential equations.  Specifically, we can show that if the initial source vertex is in the giant component, then with high probability, for all $t$ such that $0 < t < t_0$, $U(tn) = u(t)n + o(n)$ and $S(tn)=s(t)n+o(n)$.  It follows that with high probability our calculations give the correct degree distribution of the spanning tree within $o(1)$.   



\medskip {\em Power-law degree distributions.}
While the result of the previous section shows that power-law degree distributions can be observed even when none exist, 
the evidence seems overwhelming that the Internet does, in fact, have a power-law degree distribution $P(k) \sim k^{-\alpha}$.  However, as shown in~\cite{paolo,DallAsta}, traceroute sampling on graphs of this kind can underestimate the value of $\alpha$ by under-sampling the low-degree vertices relative to the high-degree ones. Here we show experimentally that the extent of this underestimate increases with the average degree of the underlying graph.  We performed experiments on both the preferential attachment model of Barab{\'a}si and Albert~\cite{pa} and the configuration model~\cite{configuration}.

\begin{figure} [htbp]
\begin{center}
\includegraphics[scale=0.45]{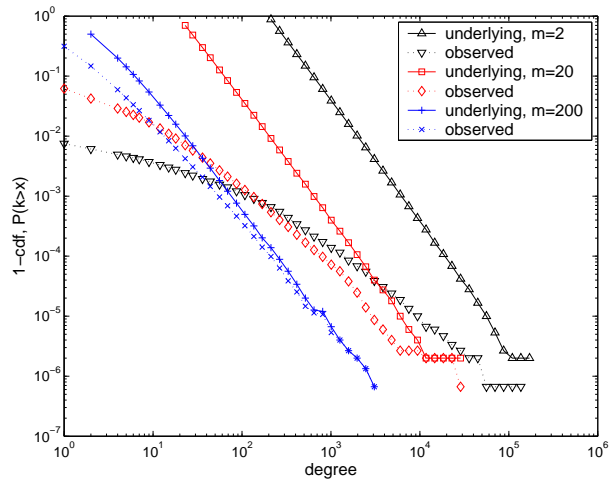}
\caption{Single-source traceroute sampling for preferential attachment networks with $n=5 \times 10^5$ and varying values of the minimum degree $m$.  The extent to which traceroute sampling underestimates $\alpha$ increases with $m$. }
\label{fig:pa}
\end{center}
\end{figure}

The preferential attachment model of~\cite{pa} gives each new vertex $m$ edges, and so has minimum degree $m$ and average degree $2m$.  In the extreme case $m=1$, the graph is a tree, and traceroutes from a single source will sample every edge.  However, as $m$ increases the fraction of edges sampled by a given source decreases.  Figure~\ref{fig:pa} shows the observed and underlying degree distributions for different values of $m$.  For $m=2$, for instance, the observed slope is $\aobs \approx 2.7$ instead of the correct value $\alpha = 3$.

It is worth pointing out that the average degree, and therefore $\aobs$, is highly sensitive to the low-degree part of the degree distribution, not just the shape of its high-degree tail.  For instance, we used the configuration model~\cite{configuration} to construct random graphs with minimum degree $k_{\min}$ and a power-law tail, i.e., $P(k)=0$ for $k < k_{\min}$ and $P(k) \propto k^{-\alpha}$ for $k \ge k_{\min}$.  (Note that the normalization of $P(k)$ then depends on $k_{\min}$.)  Here we found that $\aobs$ is a function of $k_{\min}$, not just of $\alpha$~\cite{condmatnote}.  We are currently extending our analytic calculations to this and other degree distributions.

\begin{figure} [htbp]
\begin{center}
\includegraphics[scale=0.45]{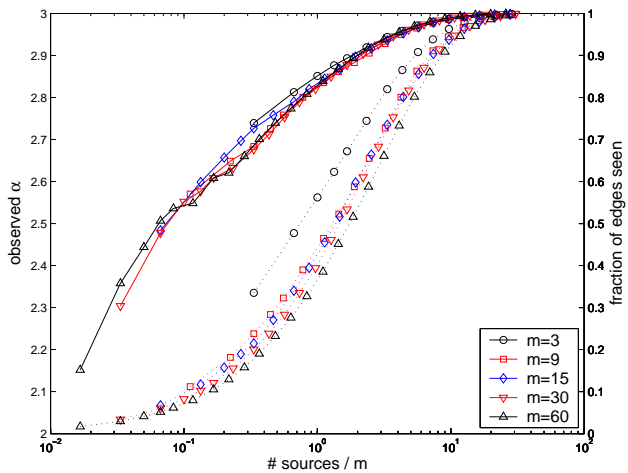}
\caption{Performance of multi-source traceroute sampling in preferential attachment networks as a function of the number of sources divided by $m$.  On the left, the convergence of $\aobs$ to the correct value $\alpha=3$; on the right, the fraction of edges observed at least once.  Both curves collapse, showing that the number of sources necessary to counter the sampling bias grows linearly with the average degree.}
\label{fig:multisource}
\end{center}
\end{figure}

\medskip {\em Building unbiased maps.}
Since single-source traceroutes can significantly underestimate $\alpha$, especially for graphs of large average degree, we now turn to the question of how many sources are needed to obtain a good estimate of $\alpha$.  In Fig.~\ref{fig:multisource}, we show the observed exponent (estimated by performing a fit to the high-degree tail $k \gg m$) for preferential attachment networks as a function of the number of sources divided by $m$; it also shows the fraction of edges included in the sample. 
The collapse of the data clearly shows that the number of sources $s$ we need to converge to within a given error from the true exponent grows linearly in $m$, and the error decreases rapidly as $s/m$ increases.  For instance, with $m$ sources we see $41\%$ of the edges and $\aobs \approx 2.82$; with $10m$ sources, $5$ times the average degree, we see $94\%$ of the edges and our estimate improves to $\aobs \approx 2.99$.   

Traceroute-based studies~\cite{Pansiot, Govindan, IMP, skitter, Opte, Faloutsos, Rocketfuel, LookingGlass, Barford} suggest an average degree for the Internet of $2.8\pm 0.5$.  (Of course, it may be higher since these studies do not see all the edges of the graph.)
However, none of these studies use more than $12$ sources, suggesting that the published values of $\alpha$ may still be somewhat low.

For the Internet, gaining access to an increasing number of sources in order to sample traceroutes from them can present practical difficulties.  However, even if the measured exponent increases with each additional source---indicating that we still do not have the correct value of $\alpha$, and the ``marginal value''~\cite{Barford} of each source is nonzero---it may be possible to extrapolate the true $\alpha$ from the rate of convergence.



\medskip {\em Conclusions.} 
Unlike the World Wide Web where links are visible, the Internet's topology must be queried indirectly, e.g., by traceroutes; and, since efficient routing protocols cause these traceroutes to approximate shortest paths, edges far from the source are difficult to see.  
Lakhina et al.~\cite{Lakhina} noted that this effect can significantly bias the observed degree distribution, and may create the appearance of a power law where none exists.  We have proved this result analytically for random graphs $G(n,p=c/n)$, showing that single-source traceroutes yield an observed distribution $P(k) \sim k^{-1}$ for $k \lesssim c$.
Other mechanisms for observing power laws in $G(n,p)$ include 
gradient-based flows~\cite{jamming}, probabilistic pruning~\cite{paolo}, and minimum weight spanning trees~\cite{barabasi}; however, these are rather different from 
our analysis.

For graphs with a power-law distribution $P(k) \sim k^{-\alpha}$ traceroute sampling underestimates $\alpha$ by under-sampling low-degree vertices~\cite{paolo,DallAsta}, and we have found that the extent of this underestimate increases with the network's average degree.  To compensate for this effect, we have found that to estimate $\alpha$ within a given error it is necessary to use a number of sources that grows linearly with the average degree.  Given the small number of sources used in existing studies, it seems possible to us that the published values of $\alpha$ for the Internet are somewhat low.  In future work, we will measure whether $\aobs$ for the Internet increases with the number of sources, and if it does, attempt to extrapolate the correct value of $\alpha$.

{\em Acknowledgments.}  
The authors are grateful to David Kempe, Mark Newman, Mark Crovella, Paolo De Los Rios, Michel Morvan, Todd Underwood, Dimitris Achlioptas, Nick Hengartner and Tracy Conrad for helpful conversations, and to ENS Lyon for their hospitality. This work was funded by NSF grant PHY-0200909 and Hewlett-Packard Gift 88425.1 to Darko Stefanovic.



\begin{thebibliography}{99}
\bibitem{BGP}
Autonomous Systems (ASs) use the Border Gateway Protocol (BGP) to route packets between networks. BGP routing tables represent logical domain adjacencies rather than physical connectivity. Tables contain advertised paths through this logical network; routing within domains is typically done via Open Shortest Path First or another distributed spanning tree protocol.
\bibitem{mult}
While there are typically multiple shortest paths from the source to a given destination, their union forms a tree if we break ties e.g.\ by giving each vertex an index and taking the first shortest path in lexicographic order.
\bibitem{condmatnote}
We have also found 
that finite-size effects can contribute to a significant over-estimate of $\alpha$ when $\alpha\le 2$. While the conjectured value for the Internet is outside this range, this effect may be relevant for other networks.
\end{thebibliography}
\end{document}